\documentclass[10pt, journal, compsoc]{IEEEtran}

\usepackage{cite, graphicx, subfigure, amsmath} 
\usepackage[]{url}
\begin{document}
%
\title{Object-oriented requirements:\\reusable, understandable, verifiable}
%
%
\author{\IEEEauthorblockN{Alexandr Naumchev}\\
\IEEEauthorblockA{Innopolis University, Innopolis, Russia\\
  University of Toulouse, Toulouse, France\\
  a.naumchev@innopolis.ru
}}
%



  \maketitle
  \begin{abstract}
  Insufficient requirements reusability, understandability and verifiability jeopardize software projects.
  Empirical studies show little success in improving these qualities separately.
  Applying object-oriented thinking to requirements leads to their unified treatment.
  An online library of reusable requirement templates implements recurring requirement structures, offering a starting point for practicing the unified approach.
  \end{abstract}


  \section{Introduction} \label{oor:intro}

  \IEEEPARstart{T}{he} industry is not actively applying requirements reuse \cite{Palomares2017}, which is regrettable: it might help, if practiced, not only to save resources in the requirements specification phase, but also to obtain documents of better quality both in content and syntax.
  It might also decrease the risk of writing low quality requirements and lead to the reuse of design, code, and tests.
  
  Bertrand Meyer in 1985 described seven understandability problems common to natural-language specifications \cite{meyer1993formalism} and proposed the process of passing them through a formal notation to produce their more understandable versions.
  He has more recently given a name to the approach -- ``The Formal Picnic Approach''\footnote{\url{https://tinyurl.com/ycn526rm}}.
  The amount of requirements and their volatility have grown, and the seven problems remain valid.
  Formal picnics should be practiced more actively and should be reusable across projects.

  The general problem of reuse finds itself in requirements' verifiability too.
  Requirements' verifiable semantics follows several recurring patterns in most of the cases \cite{Dwyer1999}.
  If a pattern exists, it should be reused, and to be reused it should be encoded as a template.
  The template should also be connected to the main instruments of software verification -- tests and contracts.
  
  Applying object-oriented thinking to the problems of requirements reusability, understandability and verifiability draws a new roadmap towards addressing them simultaneously.
  A reusable library of requirement templates, taking the familiar form of object-oriented classes, provides a starting point for practicing the approach.
  Each template encodes a formal semantics pattern \cite{Dwyer1999} as a generic class reusable across projects and components, for verifying candidate solutions through either testing or program proving.

  \section{The Problem Explained} \label{oor:the_problem_explained}

  \subsection{Reusability}\label{oor:the_problem_explained:reusability}
  Reusability has become a success story in the reuse of code \cite{Zaimi2015} and tests \cite{Tillmann2005}, but not requirements.
  On that side too, many patterns recur again and again, causing undue repetition of effort and mistakes.
  The practice of industrial projects, however, involves little reuse of requirements.
  Textual copy and subsequent modification of requirements from previous projects are still the most commonly used requirements reuse techniques \cite{Palomares2017}, which has already been long recognized as deficient in the world of code reuse.

  The most critical factors inhibiting the industrial adoption of requirements reuse through software requirement patterns (SRP) catalogues are \cite{Palomares2017}:
  \begin{itemize}
    \item The lack of a well-defined reuse method.
    \item The lack of quality and incompleteness of requirements to reuse.
    \item The lack of convenient tools and access facilities with suitable requirements classification.
  \end{itemize}

  Scientific literature studying requirements reuse approaches pays little attention to these factors when measuring the studied approaches \cite{Irshad2018}.
  The degree of reuse is the most frequently measured variable, but it is measured under the assumption that the evaluated approach is fully practiced.
  This assumption does not meet the reality: most of the practitioners who declare to practice requirements reuse approaches, apply them very selectively \cite{Palomares2017}.
  Secondary studies, which study other studies, equally ignore the factors that matter to practitioners \cite{Irshad2018}.

  \subsection{Understandability}\label{oor:the_problem_explained:understandability}
  
  Bertrand Meyer, in his work ``On Formalism in Specifications''\cite{meyer1993formalism}, described ``the seven sins of the specifier'' -- a classification of the frequently recurring flaws in requirements specifications.
  Analyzing a specification of a well-known text-processing problem illustrated that even a small and carefully written natural-language requirements document may suffer from the following problems:
  \begin{itemize}
    \item \emph{Noise} -- the presence in the text of an element that does not carry information relevant to any feature of the problem. Variants: redundancy; remorse.
    \item \emph{Silence} -- the existence of a feature of the problem that is not covered by any element of the text.
    \item \emph{Overspecification} -- the presence in the text of an element that corresponds not to a feature of the problem but to features of a possible solution.
    \item \emph{Contradiction} -- the presence in the text of two or more elements that define a feature of the system in an incompatible way.
    \item \emph{Ambiguity} -- the presence in the text of an element that makes it possible to interpret a feature of the problem in at least two different ways.
    \item \emph{Forward reference} -- the presence in the text of an element that uses features of the problem not defined until later in the text.
    \item \emph{Wishful thinking} -- the presence in the text of an element that defines a feature of the problem in such a way that a candidate solution cannot realistically be validated with respect to this feature.
  \end{itemize}
  Identified in the times when software processes were following the Waterfall model, which takes good care of every software development lifecycle phase, these problems remain.
  Nowadays processes pursue continuity, and requirements analysts have little time to process new requirements before passing them to the developers.
  The processes are iterative and collecting requirements for another iteration often starts before the current iteration finishes.
  The pace of work lowers availability of expert developers for evaluating the new requirements' verifiability.
  The pervasiveness of Internet technologies like Google Search brings problems too.
  Many sources of unclear origins now offer tons of potentially unchecked information, which is sometimes overly trusted.

  Denying the progress makes no sense, however.
  Requirements engineering tools should help the practitioners to improve the quality of information they consume and rely on.
  The improved information should be reusable across projects.

  \subsection{Verifiability}\label{oor:the_problem_explained:verifiability}

  The reusability concern applies to requirements' verifiability as well.
  Dwyer et al. analyzed 555 specifications for finite-state verification from different domains and successfully matched 511 of them against 23 known patterns \cite{Dwyer1999}.
  The patterns were encoded in modeling notations without a guidance on how to reuse them across projects for verifying candidate solutions.
  The gap still exists, and the state-of-the-practice \cite{Palomares2017} and literature reviews \cite{Irshad2018} of requirements reuse approaches, as well as the studies they cite, do not evaluate requirements' verifiability in the studied approaches.

  Requirements reuse approaches should properly address the verifiability aspect: reusing non-verifiable requirements makes little sense. The appraoches should make it clear how to capture and reuse recurring verifiable semantics' structures.

  \section{Running Example}\label{oor:running_example}

  Wikipedia represents a notable example of an intensely used and trusted Internet resource.
  The rest of the discussion relies on a Wikipedia page describing a 24-hour clock\footnote{\url{https://tinyurl.com/ybocy485}} as a requirements document example.
  The ``24-hour clock'' document is prone to the seven requirements understandability problems \cite{meyer1993formalism}.
  It only has few statements relevant to clock behavior:
  \begin{enumerate}
    \item The 24-hour clock is a way of telling the time in which the day runs from midnight to midnight and is divided into 24 hours, numbered from 0 to 24.
    \item A time in the 24-hour clock is written in the form hours:minutes (for example, 01:23), or hours:minutes:seconds (01:23:45).
    \item Numbers under 10 usually have a zero in front (called a leading zero); e.g. 09:07.
    \item Under the 24-hour clock system, the day begins at midnight, 00:00, and the last minute of the day begins at 23:59 and ends at 24:00, which is identical to 00:00 of the following day.
    \item 12:00 can only be mid-day.
    \item Midnight is called 24:00 and is used to mean the end of the day and 00:00 is used to mean the beginning of the day.
  \end{enumerate}
  The rest of the text is \emph{noise}.
  The ``or'' connective in Statement 2 results in \emph{wishful thinking}: is it acceptable to decide between the two options for every clock object, or should the decision be taken once and uniformly applied to all objects? None of the requirements after Statement 2 talk about seconds, from which it follows that the author silently made the choice in favor of the ``hours:minutes'' format.
  This ``sin'' falls into the \emph{silence} category.
  The ``usually'' qualification introduces the \textit{wishful thinking} problem to Statement 3: how are the developers expected to check candidate solutions against this requirement? Statements 4 and 6 result in a \emph{contradiction} each other: statement 4 says that midnight is 00:00, while statment 6 defines \emph{24:00} as midnight and \emph{00:00} as the beginning of the day.
  The contradiction may arise as a result of \emph{forward referencing}: \emph{24:00} and \emph{00:00} are only defined in 6, while first used in 1 and 4.
  The last part of Statement 4 is a \emph{remorse}: the author implicitly admits that the first part of the statement was not enough and adds the ``which is\dots'' part.
  Statement 5 introduces an \emph{ambiguity}, since the document never defines the ``mid-day''.
  Moreover, terms like ``mid-day'', ``midnight'', ``afternoon'' should be defined through specific clock states; it is not clear then what the author means by saying that a specific state can only be mid-day/midnight/afternoon: it can be whatever, depending on the terminology.
  
  The illustration of the object-oriented requirements approach handles a fragment of Statement 1.: ``the day runs from midnight to midnight'', referred to as ``Statement 1.1''.
  Understanding this requirement's treatment will suffice to understand the approach.
  A GitHub repository \footnote{\url{https://tinyurl.com/y6w7nlcs}} hosts the complete treatment of the ``24-hour clock'' example.

  \section{Reuse Methodology}\label{oor:reuse_methodology}

  Requirements reuse methodologies are essentially bidimensional \cite{Irshad2018}.
  The first dimension, known as \emph{development for reuse}, describes the procedure of identifying and capturing new requirement patterns.
  The second dimension, known as \emph{development with reuse}, describes the process of searching and reusing the captured patterns for specifying new requirements with lower efforts as compared to specifying them without the patterns.

  \subsection{Development for Reuse}\label{oor:reuse_methodology:dev_for_reuse}

  Given a collection of requirements:
  \begin{enumerate}
    \item Perform the standard commonality and variability analysis on the collection.
    \item Capture the identified commonality in an object-oriented class.
    \item Capture the semantical commonality through a contracted routine \cite{Tillmann2005}, \cite{Naumchev2016CompleteDrivers} to support verification.
    \item Capture the structural commonality through a string function to support formal picnics.
    \item Parameterize the identified variability points through abstraction and genericity.
  \end{enumerate}

  \subsection{Development with Reuse}\label{oor:reuse_methodology:dev_with_reuse}
  Given an informal requirement:
\begin{enumerate}
  \item Analyze the requirement's meaning and structure.
  \item Find the most appropriate requirement template class through the IDE's search facilities.
  \item Inherit from the found template in a new class representing the requirement.
  \item Refine the abstractions into domain definitions.
  \item Replace the genericity with the specified types and domain definitions.
  \item Perform a formal picnic to see if the new string representation of the requirement has a different meaning from the original one.
  \item Verify candidate solutions through running \cite{Tillmann2005} or proving \cite{Naumchev2016CompleteDrivers} the contracted routine.
\end{enumerate}

\section{Technical Artifacts}\label{oor:technical_artifacts}

Two major technical contributions support the method.

\subsection{Library of Templates}\label{oor:technical_artifacts:eiffel_library}

A ready-to-use GitHub library\footnote{\url{https://tinyurl.com/ybd4b5un}} of template classes captures known requirement patterns \cite{Dwyer1999}.
The library represents a result of applying the \emph{development for reuse} process to the patterns and provides basis for \emph{development with reuse}.
The library is written in Eiffel for readability, but the method scales to other object-oriented languages with support for genericity.

\subsection{Library of Multirequirement Patterns}\label{oor:technical_artifacts:onenote_library}

An online OneNote notebook \footnote{\url{https://1drv.ms/u/s!AsXOYPvbmuEyh4IsDdYj-i6V5yX0OA}} rearranges the original collection of patterns \footnote{\url{http://patterns.projects.cs.ksu.edu}} in the form of multirequirements  \cite{Meyer13Multi} to support their understanding.
Dwyer et al. have initially developed the patterns in 5 notations: LTL, CTL, GIL, Inca, QRE.
Their online collection consists of 5 large pages corresponding to these notations.
The alternative collection consists of 23 pages making it possible to study individual patterns in all the 5 notations simultaneously.
The representations are clickable and lead to their sources in the original repository developed by Dwyer et al.
Each page includes a link leading to the corresponding template in the GitHub library.

  \section{Applying a Template}\label{oor:applying_template}
  The following illustration handles the ``Statement 1.1'' requirement by applying a reusable template class from the GitHub library.
  The requirement fits into the ``Global Response'' pattern \cite{Dwyer1999}.
  The pattern reads: ``S responds to P globally'', for events S and P.
  It is the most frequently used pattern: out of the 555 analyzed requirements \cite{Dwyer1999}, 241 represented this pattern.
  For ``Statement 1.1'', both S and P map to the midnight event: ``midnight responds to midnight globally''.
  This new statement paraphrases the original one, ``the day runs from midnight to midnight''.
  \begin{figure*}[t]
    \centering
    \subfigure[EiffelStudio with the \emph{STATEMENT\_1\_1} class representing the ``Statement 1.1'' requirement.]
    {
      \includegraphics[width=\textwidth]{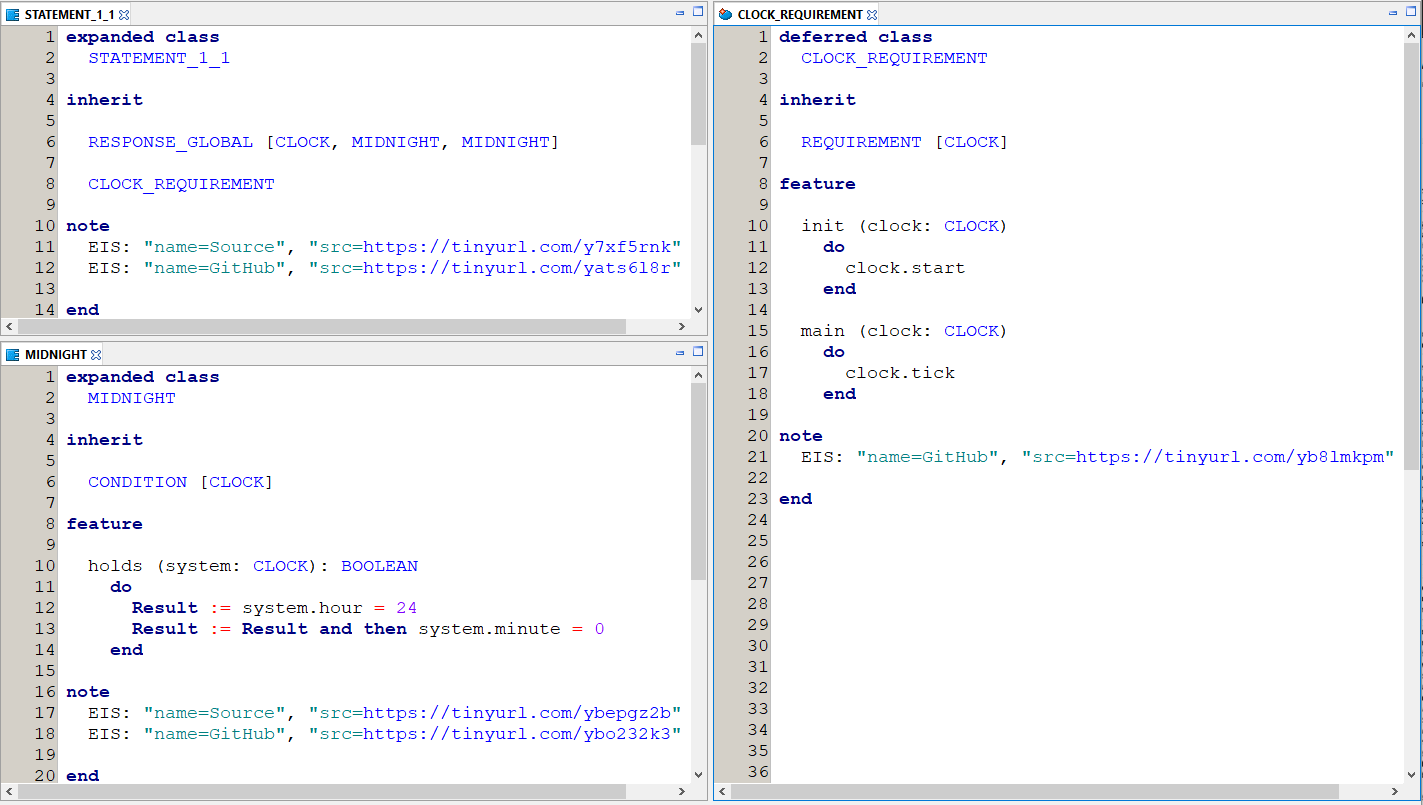}
      \label{fig:oor:env:es}
    }
    \subfigure[Google document with the contents of the ``24-hour clock'' Wikipedia page.]
    {
      \includegraphics[width=\textwidth]{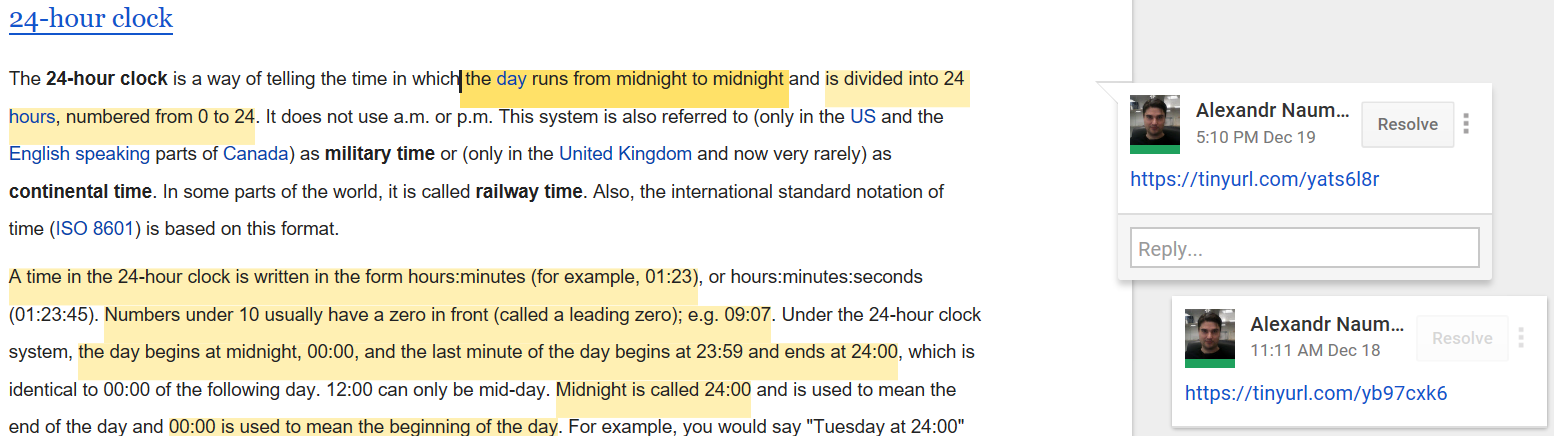}
      \label{fig:oor:env:gd}
    }
    \caption{Requirement classes in EiffelStudio (Figure \ref{fig:oor:env:es}), and the contents of the ``24-hour clock'' Wikipedia page copied to a Google document (Figure \ref{fig:oor:env:gd}). The ``Source'' link in the \emph{STATEMENT\_1\_1} class leads to the corresponding commented fragment in the Google document. The comment contains the GitHub location of the fragment's object-oriented version, equal to the location in the ``GitHub'' EIS link in \emph{STATEMENT\_1\_1}.}
    \label{fig:oor:env}
  \end{figure*}

  Class \emph{STATEMENT\_1\_1} (Figure \ref{fig:oor:env:es}) captures the requirement.
  The class inherits from:
  \begin{itemize}
    \item A generic application of class \emph{RESPONSE\_GLOBAL} to classes \emph{CLOCK} and \emph{MIDNIGHT}, where \emph{RESPONSE\_GLOBAL} is a generic template encoding the ``Global Response'' pattern.
      The \emph{RESPONSE\_GLOBAL [CLOCK, MIDNIGHT, MIDNIGHT]} application reads: ``for type \emph{CLOCK}, \emph{MIDNIGHT} response to \emph{MIDNIGHT} globally''.
    \item Class \emph{CLOCK\_REQUIREMENT} recording domain information common to all clock requirements: the fact that the \emph{tick} routine advances a clock's state, and the \emph{start} routine initializes a new clock.
  \end{itemize}
 
  The \emph{CLOCK} class is a candidate solution implementing the ``clock'' concept, and the \emph{MIDNIGHT} class captures the definition of midnight through effecting the deferred \emph{holds} Boolean function inherited from generic class \emph{CONDITION} applied to the \emph{CLOCK} class.
  The generic application emphasizes the fact that the notion of midnight applies to the notion of clock.

  The classes have something in common: the ``note'' section at the bottom with Web links of two kinds.
  Links named ``Source'', when followed, highlight the fragments in the original requirements documents from which the enclosing requirement classes were derived.
  Links named ``GitHub'', when followed, lead to the enclosing classes' locations on GitHub.
  The ``Source'' link in \emph{STATEMENT\_1\_1}, for example, highlights, when followed, the ``the day runs from midnight to midnight'' phrase in the Google document\footnote{\url{https://tinyurl.com/y96rj2v3}}, and brings the comment on this phrase to the reader's attention (Figure \ref{fig:oor:env:gd}).
  The comment contains the GitHub link leading back to the \emph{STATEMENT\_1\_1} class on GitHub; this link is identical to the ``GitHub'' link in the \emph{STATEMENT\_1\_1} class' ``note'' section.

  \section{Formal Picnic}\label{oor:formal_picnic}

  The \emph{RESPONSE\_GLOBAL} class implements its string representation through redefining the standard \emph{out} function present in all Eiffel classes.
  Any instruction that expects a string argument, such as \emph{print}, automatically invokes this function to get the argument's string representation if the argument has a non-string type.
  \begin{figure*}[t]
    \centering
    \includegraphics[width=0.6\textwidth]{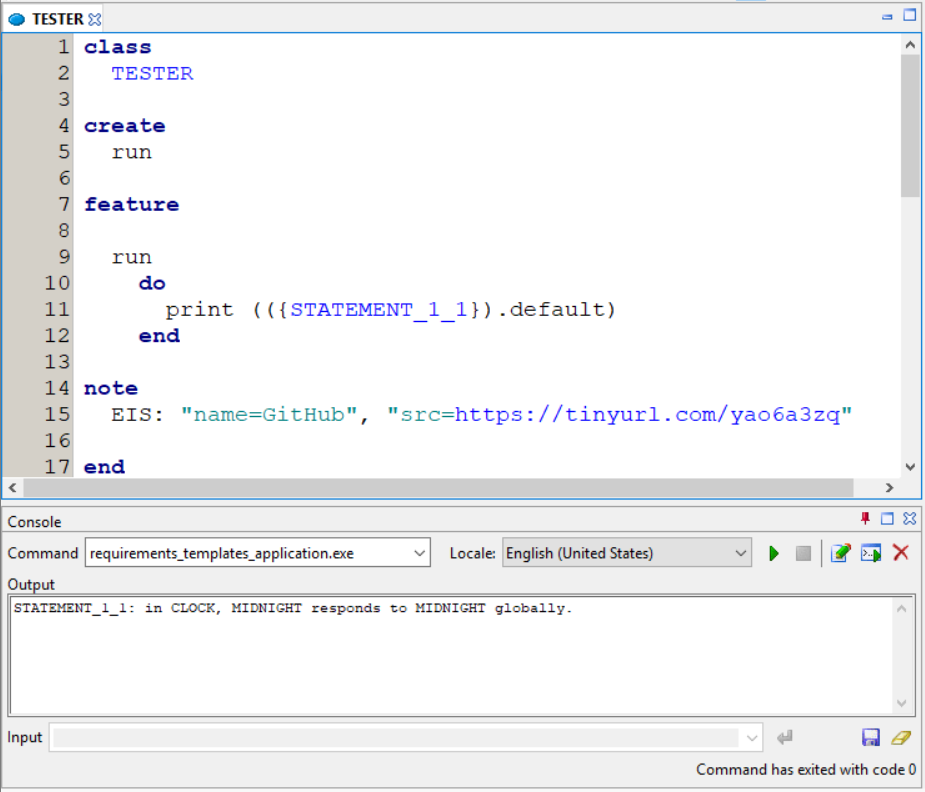}
    \caption{The executable code (the upper window) outputs the automaticaly generated string representation of the requirement to the console (the lower window).}
    \label{fig:oor:fp}
  \end{figure*}

  Routine \emph{run} of class \emph{TESTER} (Figure \ref{oor:formal_picnic}) is a configurable entry point of the console application illustrating formal picnics and verification of object-oriented requirements.  

  Line 11 of \emph{TESTER} outputs the structured string representation of the \emph{STATEMENT\_1\_1} object-oriented requirement.
  The \emph{.default} expression returns the default object of the \emph{STATEMENT\_1\_1} class, and the \emph{print} instruction puts the object's string representation to the ``Output'' window below the ``TESTER'' window.
  The requirement's name, ``STATEMENT\_1\_1'', goes before the colon and its string representation goes after.

  The requirements analyst now has two comparable string representations of the requirement: the original and the generated one.
  Comparing them facilitates analysis and may result in asking clarifying questions to the customer and in additional communication.

  \section{Verification}\label{oor:verification}

  The template classes, including \emph{RESPONSE\_GLOBAL}, contain instruments of their own verification in the form of a contracted routine called ``verify''.
  The \emph{run} routine of the \emph{TESTER} class may call \emph{verify} to test a candidate solution.

  \begin{figure*}[t]
    \centering
    \includegraphics[width=\textwidth]{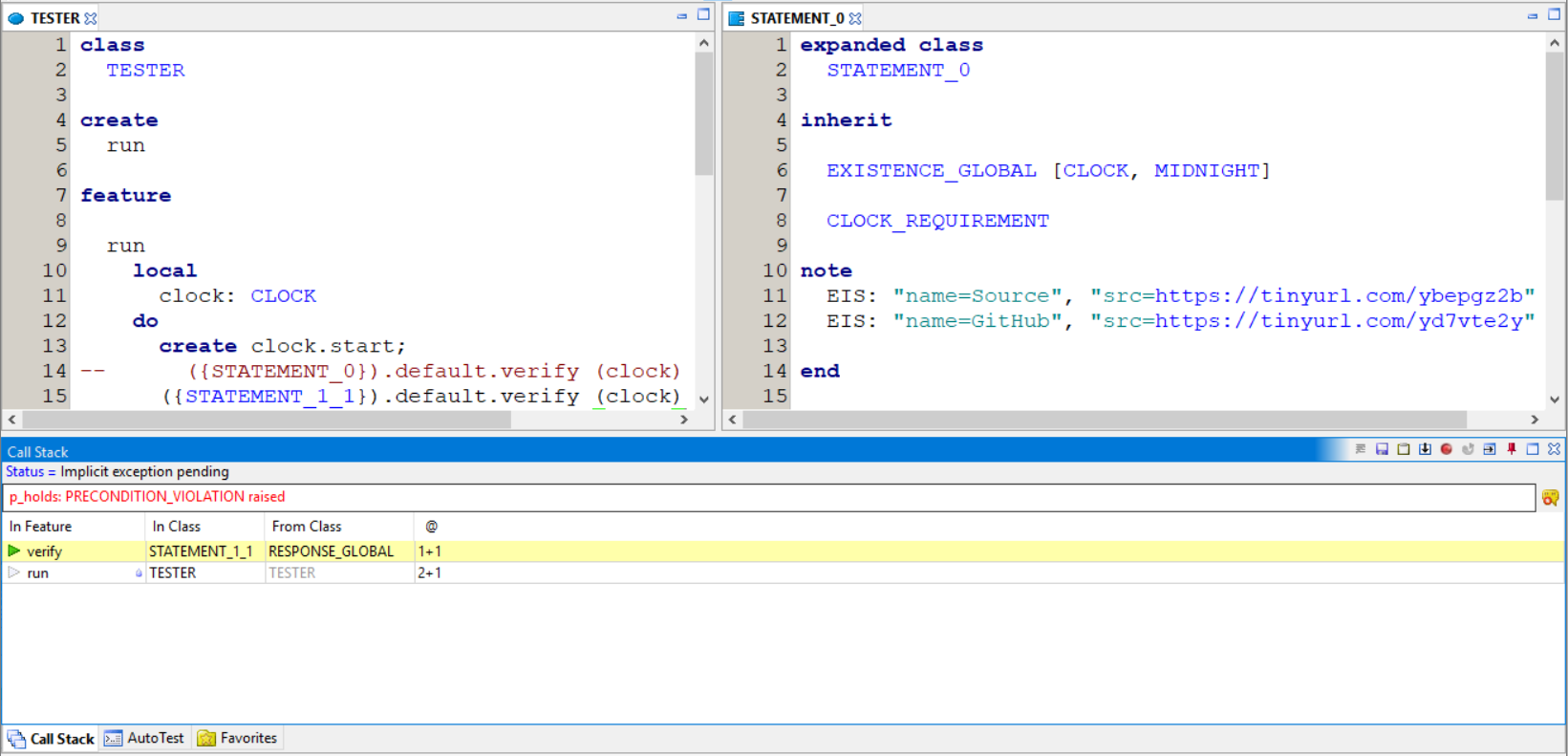}
    \caption{An exception caused by violating the requirement's verification precondition.}
    \label{fig:oor:verification}
  \end{figure*}
  
  Line 15 of the \emph{TESTER} class (Figure \ref{fig:oor:verification}) tests class \emph{CLOCK} as a candidate solution of the \emph{STATEMENT\_1\_1} requirement.
  Line 13 instantiates a \emph{CLOCK} variable, while lines 14 and 15 use the variable as test input.
  The following discussion explains the nature of line 14.
  The line is commented to illustrate the problem that the line fixes when uncommented.
  
  The \emph{verify} routine has a precondition.
  For the \emph{STATEMENT\_1\_1} class, the precondition becomes the \emph{holds} Boolean function from the \emph{MIDNIGHT} class.
  This function returns \emph{True} only for the \emph{24:00} time, and the newly instantiated \emph{clock} variable is set to time \emph{00:00}.
  Line 14 fixes this mismatch, and its removal crashes the execution.
  The ``Call Stack'' window provides information related to the failure: a precondition tagged ``p\_holds'' is violated in \emph{STATEMENT\_1\_1}, inherited from the \emph{RESPONSE\_GLOBAL} template class.
  The testing code should set the \emph{clock} variable's state to time \emph{24:00} before testing \emph{STATEMENT\_1\_1}; line 14 does exactly this.
  \emph{STATEMENT\_0} is a requirement class saying that the midnight state should be in principle achievable by \emph{CLOCK}.
  The \emph{EXISTENCE\_GLOBAL} pattern \cite{Dwyer1999} captures this semantics.
  Line 14 tests \emph{CLOCK} against \emph{STATEMENT\_0} by trying to reach the midnight state on the input variable.
  Uncommenting the line will remove the precondition violation.

  The process of deriving \emph{STATEMENT\_0} is an example of how the verification process may help identify a new requirement and learn a new template.

  Program proving and Design by Contract may be used instead of testing.
  The automatic prover (AutoProof \cite{tschannen2015autoproof} in the context of Eiffel) should be applied to the requirements classes, \emph{STATEMENT\_0} and \emph{STATEMENT\_1\_1}.
  The prover will statically check the contracted \emph{verify} routine according to the principles of Hoare logic \cite{hoare1969axiomatic}.
  The prover will only accept the routine if the \emph{CLOCK} class has a strong enough and correct contract \cite{Naumchev2016CompleteDrivers}.
  The illustration relies on testing because AutoProof, in its current state, requires a lot of additional annotations to check classes like \emph{STATEMENT\_1\_1}, and explaining these annotations goes beyond the object-oriented requirements idea's essentials.

  \section{Assessment}\label{oor:assessment}

  The approach helps to fix the identified problems undermining the lack of requirements reuse:
  \begin{itemize}
    \item \emph{The lack of a well-defined reuse method}: the reuse method is object-oriented software construction, which is a well-defined method.
    \item \emph{The lack of quality and incompleteness of requirements to reuse}: the templates library implements the existing collection of specification patterns proven to cover most of the cases, which makes the library complete and quality in that sense.
    \item \emph{The lack of convenient tools and access facilities with suitable requirements classification}: the tools and access facilities are object-oriented IDEs and GitHub, with all their powerful features.
  The classification is that of the Dwyer et al.'s collection, proven to be practically relevant.
  \end{itemize}

  The approach helps to fix the requirements understandability problems:
  \begin{itemize}
    \item \emph{Noise}: only those requirements remain that fall into an existing verifiable requirement template.
    \item \emph{Silence}: an attempt to verify existing object-oriented requirements may uncover missing requirements, as it was the case with \emph{STATEMENT\_0}.
    \item \emph{Overspecification}: only those requirements remain that fall into an existing verifiable requirement template.
  Implementation details cannot map to a requirement template.
    \item \emph{Contradiction}: one notion may be defined in only one way, otherwise the IDE will raise a compilation error.
      The contradiction caused by two inconsistent definitions of midnight was resolved by defining this notion in the form of the \emph{MIDNIGHT} class.
    \item \emph{Ambiguity}: little can be done to remove the possibility for different interpretations -- the requirements interpretation process is performed by a cognitive agent anyway.
  If an interpretation is identified as erroneous, however, switching to another template will automatically update both the generated string representation and the underlying verifiable semantics.
  In other words, the templates may help to reduce the effort spent on fixing the consequences of the misinterpretation.
    \item \emph{Forward reference}: the approach removes this problem.
  There is no notion of requirements' order in the object-oriented approach, and meaningful statements are connected by the standard ``client-supplier'' relationship, extensively supported by the object-oriented IDEs.
    \item \emph{Wishful thinking}: only those requirements remain that fall into an existing verifiable requirement template.
  The compiler will not accept a template's application in which the verifiable semantics is not fully defined.
  \end{itemize}
  
  The approach helps to fix the requirements verifiability problem.
  The GitHub library of classes fixes the lack of reusable templates covering the identified verifiable specification patterns.
  The approach makes it possible to capture and reuse newly identified patterns using the existing object-oriented techniques complemented with contracts. 

  Besides the benefits, the approach has some limitations:
  \begin{itemize}
    \item Requirements analysts' familiarity with the principles of object-oriented analysis and design.
    \item Software developers' familiarity with the principles of Hoare logic based reasoning.
  \end{itemize}

  \section{Supporting Work}\label{oor:supporting_work}

  The idea to use a programming language as a requirements notation is not new \cite{Meyer13Multi}, \cite{Naumchev2016UnifyingExample}, \cite{Naumchev2017}, \cite{DBLP:journals/corr/abs-1710-02801} and is well justified.
  Many groups of stakeholders prefer descriptions of operational activity paths over declarative requirements specifications \cite{Sindre2003}.
  A demand exists for educating developers capable of both abstracting in a problem space and automating the transition to a solution space \cite{Whittle2014}.

  Other approaches to requirements reuse do not share the aspirations towards connecting the requirements and the solution spaces, as follows both from the state-of-the-practice \cite{Palomares2017} and the literature \cite{Irshad2018} studies.
  The studied approaches focus on reusing natural language, use cases, domain models and several other artifacts disjoint from the solution space.

  The decision to express requirements in a programming language may bridge the gap.
  It may also be the only way to bring the developers closer to the requirements they implement: industry practitioners are generally not keen to switching their tools \cite{Dalpiaz2018}.
  The advanced state of code reuse has all chances to skyrocket the state of requirements reuse if the requirements take the form of code.

  \section{Future Work}\label{oor:future_work}

  Intelligent tools should be embedded into existing text editors for:
\begin{itemize}
  \item Detecting known patterns in what requirements analysts specify manually.
  \item Proposing reusable templates corresponding to the identified patterns.
  \item Identifying new patterns in requirements that do not map to existing patterns.
\end{itemize}
Natural language processing (NLP) would be an appropriate instrument for implementing these tools \cite{Dalpiaz2018}.



  \bibliographystyle{IEEEtran}
  \bibliography{IEEEabrv, library}

  %
%



  \end{document}